%% file: main.tex
\documentclass[trackchanges,twocolumn]{aastex701}
\newcommand{\msun}{M$_{\odot}$}

\input{macros}
\newcommand{\grs}{$\gamma$ rays}
\newcommand{\gr}{$\gamma$-ray}
\usepackage{enumerate}
\usepackage{tabularx}
\usepackage[flushleft]{threeparttable}
\usepackage{amsmath}

\begin{document}

\title{Discovery and Timing of the First Millisecond Pulsar in NGC~6316}

\correspondingauthor{D. R. Bhakta}

\author[orcid=0000-0002-7965-3076,gname=Deven,sname=Bhakta]{D. R. Bhakta}
\affiliation{Department of Astronomy, University of Virginia, 530 McCormick Road, Charlottesville, VA 22904, USA}
\affiliation{National Radio Astronomy Observatory, 520 Edgemont Road, Charlottesville, VA 22903, USA}
\email[show]{drb7tg@virginia.edu}  

\author[orcid=0000-0001-5799-9714,gname=Scott,sname='M. Ransom']{S. Ransom}
\affiliation{National Radio Astronomy Observatory, 520 Edgemont Road, Charlottesville, VA 22903, USA}
\email{sransom@nrao.edu}

\author[orcid=0000-0002-2185-1790,gname=Megan,sname=DeCesar]{M. DeCesar}
\affiliation{College of Science, George Mason University, Fairfax, VA 22030, USA}
\email{}

\author[orcid=0000-0002-9618-2499,gname=Shi,sname=Dai]{S. Dai}
\affiliation{Australia Telescope National Facility, CSIRO, Space and Astronomy, PO Box 76, Epping, NSW 1710, Australia}
\affiliation{Western Sydney University, Locked Bag 1797, Penrith South DC, NSW 2751, Australia}
\email{}

\begin{abstract}
NGC~6316 is a poorly studied, distant, and massive globular cluster (GC) with prominent gamma-ray emission detected via the \textit{Fermi} Large Area Telescope (LAT). Based on gamma-ray spectral studies, NGC~6316 is expected to host tens of millisecond pulsars (MSPs). Using the Green Bank Telescope (GBT) and Murriyang, CSIRO’s Parkes radio telescope (Parkes), we present the discovery and a 3.1\,yr duration timing solution of the first millisecond pulsar found in the cluster. PSR~J1716$-$2808A has a rotational period of 2.45\,ms and is in a binary with a $\sim$0.1\,\msun \ companion with an orbital period of 0.42\,d. This is a typical cluster MSP within a compact orbit with no evidence of eclipses. PSR~J1716$-$2808A has a dispersion measure DM = 172.26\,pc\,cm$^{-3}$, which is lower than predicted NE2001, YMW16 and NE2025 electron density model values. The MSP is located within half a core radius from the cluster center and has a negative period derivative, implying that it is on the back side of the cluster and is being accelerated towards us. Given the negative period derivative, we report an upper limit on the maximum line-of-sight cluster acceleration, $a_{l,\textrm{GC}}/c \approx -2.3\times10^{-18}$\,s$^{-1}$, experienced by the pulsar and constraints on the magnetic field to be $<\sim$3$\times$10$^{8}$\,G. The presence of external acceleration strongly supports the pulsar to be within NGC~6316. We can better constrain NGC~6316's properties through longer-term timing of PSR~J1716$-$2808A or by finding more pulsars within the cluster. Based on the gamma-ray pulsar estimates and a cluster distance of 11.3\,kpc, deeper, more sensitive searches would find many additional pulsars. 
\end{abstract}


\section{Introduction}
Millisecond pulsars are created preferentially in globular clusters (GCs) when compared, per unit stellar mass, to the Galactic disk. In contrast with pulsars in the Galactic field, of which $\sim$ 10\% are in binary systems \citep{manchester_australia_2005} \footnote{\url{http://www.atnf.csiro.au/research/pulsar/psrcat}}, more than 50\% of all known GC pulsars are in binary systems\footnote{See \url{https://www3.mpifr-bonn.mpg.de/staff/pfreire/GCpsr.html} for a list ++of all known GC pulsars}. This is due to the large stellar density and the resulting interactions that occur within the clusters \citep{fru00,ver14,phi93,wu21}. Searching for and conducting long-term timing of these MSPs can yield significant insight into these GCs. Multiple cluster MSPs can be timed to constrain and model the cluster's ionized gas content \citep{ran08, ye19}. Short-period eclipsing binaries can provide information on pulsar winds, and the nature of the companion stars \citep{fre05bin}. Furthermore, binary parameters can be measured and used to study binary evolution, and subsequent NS mass measurements can constrain the nuclear equation of state \citep{zha11,oze16,tha20}. The period derivatives and position measurements of these cluster pulsars can provide measurements on the host cluster's mass distribution and dynamics due to the period derivatives being dominated by gravitational accelerations from the cluster \citep{phi93,ran08, pra17}.

\textit{Fermi}-LAT has detected \grs\ from multiple GCs, likely coming from the integrated emission from dozens of MSPs in the clusters \citep{abd09,abd10,abd20,son21,fre11,wu13}. Gamma-ray pulsations from MSPs within several clusters have been detected, albeit rare; only 3 GC MSPs have been confirmed as \gr \ MSPs: J1823$-$3021A in NGC~6624 \citep{fre11}, B1821$-$24 in NGC~6626 \citep{wu13}, and J1835$-$3259B in NGC~6652 \citep{zhang_discovery_2022}, with a 4th likely detection from J1717+4308A in NGC~6341 \citep{zhang_likely_2023}. The rarity of \gr\ GC MSP detections is likely due to high background and source confusion due to the crowded fields. Estimates based off of the \gr \ spectral properties indicate tens of pulsars should populate several of the brightest clusters \citep{dem19,wu21}, including NGC~6316. 

NGC~6316 is a scarcely investigated massive (M $\sim$ 3.5$\times$10$^{5}$ M$_{\odot}$, top 20\% of GCs) bulge cluster located approximately 11.3\,kpc \citep{der23} from the Sun \citep{bau19}. \cite{dav92} and more recently, \cite{der23} noted similarities between NGC~6316 and 47~Tuc based on their color-magnitude diagrams and age. 47~Tuc currently has 42 known pulsars within the cluster (behind Terzan~5's 49 pulsars)\footnote{\url{https://www3.mpifr-bonn.mpg.de/staff/pfreire/GCpsr.html}}. Studies of the \gr \ emission predict anywhere from 13 MSPs \citep{wu21} to 10$^{2}$ MSPs \citep{dem19} within NGC~6316. Recent modeling extrapolating new FAST discoveries with the current GC population predicts between 6 and 20 pulsars within NGC~6316 \citep{yin_fast_2024}. We have conducted radio searches of NGC~6316 alongside several other \gr\ bright or massive globular clusters to search for new MSPs. As a part of this search, we have observed the GCs: NGC~6304, NGC~6717, NGC~6139, NGC~6388, GLIMPSE-C02, 2MASS-GC01, Terzan 2, and Liller 1, with data reduction and searches on-going for the other clusters. In this paper, we report the discovery and timing of a highly accelerated MSP within a compact binary system in NGC~6316.

\begin{table*}[]
\centering
\begin{tabular}{|c|c|c|c|c|c|c|c|c|c|}
\hline
\textbf{Observatory} & \textbf{Receiver} & \textbf{Dedispersion} & \textbf{Central Freq.} & \textbf{Bandwidth} & \textbf{T$_{obs}$}  & \textbf{T$_{int}$} & \textbf{N$_{chan}$} & \textbf{N$_{obs}$} \\
& & (pc\,cm$^{-3}$) & (MHz) & (MHz) & (hr) & ($\mu$s) & & \\
\hline
GBT & C-band & Incoherent & 6012 & 4500 & 2 & 43.69 & 1024 & 2\\
GBT & S-band & Incoherent & 2165 & 1500 & 2 & 10.92 & 1024 & 1\\
GBT$^a$ & S-band & Coherent (281.5) & 2165 & 1500 & 2 & 43.69 & 1024 & 1\\
Parkes & UWL & Coherent (172) & 2368 & 3328 & 3, 3.5 & 64 & 3328 & 18\\
GBT & L-band & Coherent (172) & 1500 & 800 & 5 & 10.24 & 512 & 2\\ \hline
\end{tabular}
\caption{Observation set up parameters for all NGC~6316 observations. All data was recorded in 8-bit format. All observations were recorded in total intensity, except for GBT L$-$band which was full stokes. Our Parkes observations were a combination of 3 and 3.5\,hr scheduling blocks. \textbf{$^a$}The initial detection of PSR~J1716$-$2808A. \vspace{-5mm}}
\label{obs_setup}
\end{table*}

\section{GBT Observations}
\begin{figure*}[]
    \centering
    \includegraphics[width=0.9\linewidth]{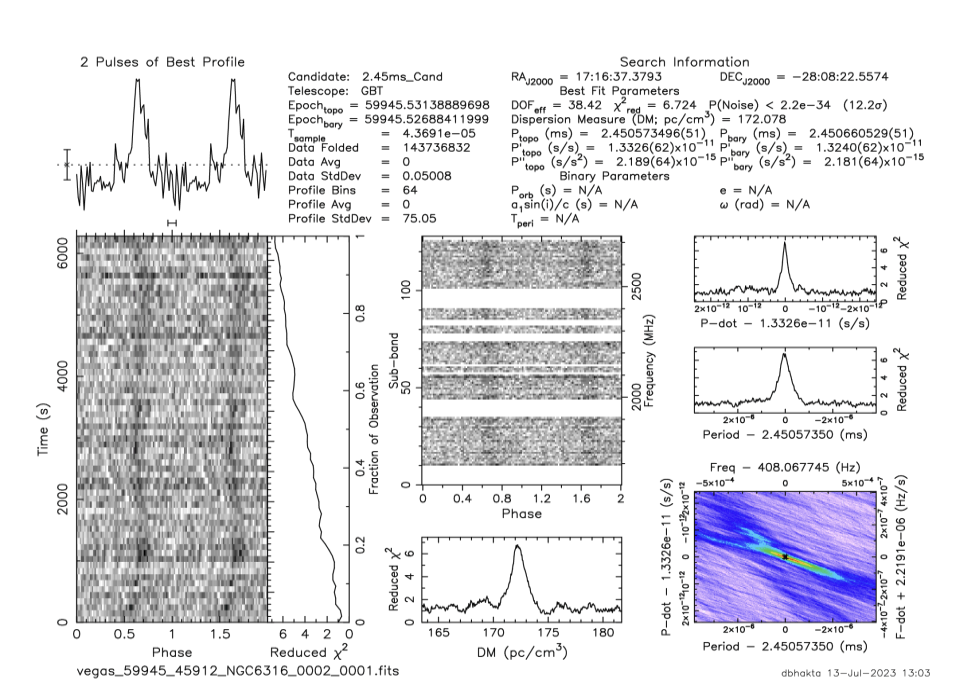}
    \caption{The {\tt prepfold} output of the first detection of PSR~J1716$-$2808A from a GBT S$-$band observation. The residual curvature in the time vs pulse phase greyscale on the left indicates uncorrected orbital motion despite fitting for acceleration and jerk (i.e.~$\dot p$ and $\Ddot{p}$).}
    \label{initial}
\end{figure*}

We observed NGC~6316 with the Robert C. Byrd Green Bank Telescope (GBT; \cite{pres09}) using the C$-$band and S$-$band receivers, and observed late August 2022 and late December 2022 respectively (project ID: AGBT22B\_270, PI: D. Bhakta). The observation parameters are listed in Table.~\ref{obs_setup}. 

We used {\tt PRESTO}\footnote{\url{http://www.cv.nrao.edu/~sransom/presto/}} \citep{presto} to search for periodic pulsations in the GBT S$-$band and C$-$band data with {\tt rfifind} to mask and zap radio-frequency interference. We dedispersed the data into barycentric time series ranging 0-471\,pc\,cm$^{-3}$ using {\tt prepsubband} and {\tt DDplan.py} with 471\,pc\,cm$^{-3}$ selected as the max predicted DM via the NE2001 and YMW16 electron density models (371\,pc\,cm$^{-3}$) $+$100\,pc\,cm$^{-3}$ to account for uncertainties in DM estimates \citep{ne2001,ymw16}. In order to search for short orbital period pulsars, we searched the observations in 30-45 minute durations in addition to the full 2\,hr duration. We used the routine {\tt accelsearch} to search for periodic signals in the Fourier domain, with $z_{max} = $ 10, 200 and 600 for the full duration and $z_{max} = $ 200 for the shorter duration searches. $z_{max}$ represents the maximum number of Fourier frequency bins that a signal can drift during an individual observation due to a pulsar's acceleration \citep{rem02}. Using {\tt ACCEL\_sift.py}, we filtered and identified candidates for folding with {\tt prepfold} within {\tt PRESTO}.

\section{PSR~J1716$-$2808A detection}
We identified a highly accelerated pulsar candidate in the $z_{max} =$ 600 full duration search (z = 146.0 corresponding to an acceleration of 2.7\,m\,s$^{-2}$) with a spin period $P \simeq$ 2.45\,ms at DM $\simeq$ 172\,pc\,cm$^{-3}$ in the second S$-$band observation (coherently dedispersed at DM 281.5\,pc\,cm$^{-3}$) on MJD 59945. Due to the highly accelerated nature of the candidate, we searched over $\Ddot{P}$ with {\tt prepfold} in order to fold the new MSP through the full observation shown in Fig~\ref{initial}. We detected the new MSP, PSR~J1716$-$2808A in the incoherently dedispersed S$-$band observation from MJD 59939 but not the C$-$band observations. The large acceleration indicated that this MSP is in a compact binary. We conducted a DDT observation with Murriyang, CSIRO’s Parkes radio telescope (hereafter “Parkes”) to check if the MSP is detectable with the Parkes ultra-wide-bandwidth, low frequency receiver (UWL) \citep{hobbs_ultra-wide_2020} before following up with a dedicated timing campaign. 

\subsection{Parkes Timing}
Our Parkes DDT observation occured on MJD 60282 for 3.5\,hrs. The observing setup used the {\tt MEDUSA} backend with 3328\,MHz observing band across 26 subbands (128\,MHz per subband), 64\,us sampling, and coherent dedispersion at DM 172.00\,pc\,cm$^{-3}$ in search mode. We processed and analyzed the data identically to the GBT observations with no detection using acceleration searches over the full observation as well as with short duration searches. Given the two S$-$band detections, we created a family of circular orbital solutions using {\tt fit\_circular\_orbit.py} from {\tt PRESTO}. These possible orbits were applied to topocentric timeseries generated at the DM of the S$-$band detections to search for a detection using {\tt SPIDER\_TWISTER\footnote{\url{https://alex88ridolfi.altervista.org/pagine/pulsar_software_SPIDER_TWISTER.html}}}.

\begin{figure}
    \centering
    \includegraphics[width=\linewidth]{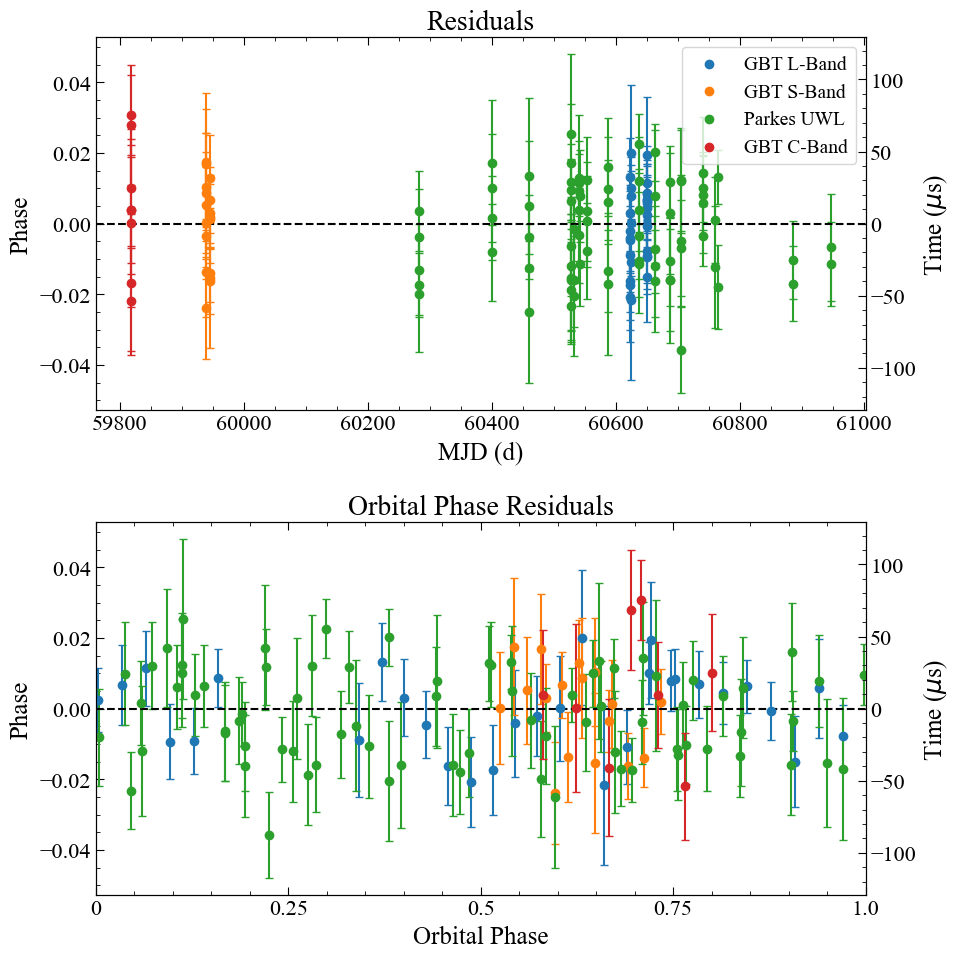}
    \caption{Top: Time and Phase Residuals of PSR~J1716$-$2808A over a 3\,yr timing baseline. The central observing frequencies for each observation are reported in Table.~\ref{obs_setup}. Bottom: Orbital phase coverage across all of our observations shows complete coverage of the 10.2\,hr orbit. Note that eclipses, if present, would occur at orbital phase 0.25.}
    \label{resids}
\end{figure}

{\tt SPIDER\_TWISTER} folds the timeseries across many trials, where the time of passage at the ascending node ($T_{\rm asc}$) is slightly altered to search over the orbital phase of a binary, and the highest signal-to-noise value is returned. Using {\tt SPIDER\_TWISTER}, we detected PSR~J1716$-$2808A in the Parkes DDT observation. However, we were unable to further constrain the orbital solutions, as the Parkes detection overlapped with the previous detections in orbital phase. Given the successful detection with the Parkes DDT, we conducted a dedicated timing campaign using Parkes (project ID: P1330, PI: D. Bhakta) to fully solve the orbit of the binary and determine a phase-connected timing solution. Additionally, we conducted 2$\times$5\,hr observations (observed on MJDs 60623 and 60650) with the GBT at L$-$band for deeper, more sensitive pulsation searches (project AGBT24B\-430, PI D. Bhakta) with the additional advantage of the longer observing tracks providing greater orbital coverage. With the new GBT and Parkes observations, we solved the orbit and phase-connect across a 3\,yr timing baseline, see Fig.~\ref{resids}. 

\section{Results}
Timing parameters of PSR~J1716$-$2808A are reported in Table~\ref{params}. We measured astrometric position, spin frequency and spin frequency derivative, and the Keplerian orbital parameters. An upper limit on the eccentricity for the circular orbit is $e < 7\times 10^{-5}$, with the minimum companion mass of $\sim$ 0.1\,\msun. The pulsar position is $\sim$ 4.5$\arcsec$ from the cluster center and within the cluster core \citep[$r_{c} \sim 10$\arcsec,][]{der23}. PSR~J1716$-$2808A's dispersion measure is 172.26\,pc\,cm$^{-3}$, which is lower than predicted values at a distance of 11.3\,kpc from the NE2001 (DM$\sim$374\,pc\,cm$^{-3}$), YMW16 (DM$\sim$193\,pc\,cm$^{-3}$), and the recent NE2025 (DM$\sim$210\,pc\,cm$^{-3}$) electron density models \citep{ne2001,ymw16,ocker_ne2025_2026}. This is unsurprising when comparing the known DM of bulge clusters to the predicted electron density model DMs, we find that the DM of more than 50\% of bulge clusters are overestimated (by $>10\%$) in the NE2001 and YMW16 models and $\sim$40\% in the NE2025 models.

Our TOAs show full orbital coverage with no evidence of eclipses. We are not aware of any non-eclipsing redback systems, almost all of which show extended eclipses of up to $\sim$0.5 of the orbital period, and the minimum companion mass estimate is in between typical mass ranges for redbacks and black widows. While we cannot rule out the possibility of this pulsar being a black widow, it does appear to be a typical MSP in a highly accelerated, compact orbit binary system. Based on a comparison to the orbital period to the projected semi-major axis of the Galactic binary pulsar population, the companion would likely be a Helium white dwarf or an ultralight companion \citep[see Fig.~6][]{bernadich_joint_2026}. JWST observations might help determine the nature of the companion. Given the positive spin frequency derivative, the intrinsic \textit{$\dot{f}$} is likely dominated by accelerations due to the environment \citep{phi92, phi93}. 

\begin{table}
\resizebox{\linewidth}{!}{
\begin{threeparttable}
\caption{Parameters for PSR J1716$-$2808A}
\label{params}
\begin{tabular}{ll}
\hline\hline
\multicolumn{2}{c}{Dataset and model summary}\\ 
\hline
Pulsar name                  \dotfill & J1716$-$2808A      \\ 
MJD range                    \dotfill & 59818---60946 \\ 
Data span (yr)               \dotfill & 3.09    \\ 
Number of TOAs               \dotfill & 135      \\ 
Solar system ephemeris       \dotfill & DE440      \\ 
Timescale                    \dotfill & TT(BIPM2023)      \\ 
PEPOCH, Spin Reference ($\mathrm{d}$)\dotfill &  60623.685022 \\ 
Time unit                    \dotfill & TDB      \\  
Binary model               \dotfill & ELL1      \\ 
Number of EFACs          \dotfill & 5     \\ 
Number of EQUADs          \dotfill & 1      \\ 
\hline
\multicolumn{2}{c}{Measured Quantities} \\ 
\hline
Right Ascension (RA, J2000) \dotfill & $17^{\rm h}\;16^{\rm m}\;37\fs54045(19)$ \\
Declination     (DEC, J2000) \dotfill & $-28\degr\;08\arcmin\;22\farcs900(29)$ \\
Spin-frequency, $f$ ($\mathrm{Hz}$)\dotfill &  $408.0317456967(1)$ \\ 
$\dot f$ ($\mathrm{Hz\,s^{-1}}$)\dotfill &  $1.396(5) \times 10^{-15}$ \\ 
Dispersion measure ($\mathrm{pc\,cm^{-3}}$)\dotfill &  $172.265(8)$ \\ 
Orbital period ($\mathrm{d}$)\dotfill &  $0.424976623(2)$ \\ 
Projected semi-major axis ($\mathrm{ls}$)\dotfill &  $0.339253(4)$ \\ 
Time of ascending node ($\mathrm{d}$)\dotfill &  $59939.333399(4)$ \\ 
\hline
\multicolumn{2}{c}{Fit Statistics} \\ 
\hline
Reduced chi-squared value \dotfill &  1.12 \\ 
TOA residual ($\mathrm{\mu s}$)\dotfill &  30.94 \\ 
\hline
\multicolumn{2}{c}{Derived Quantities} \\ 
\hline
$p$, Spin-period ($\mathrm{s}$)\dotfill & $0.0024507897009153(7)$ \\
$\dot p$, Spin-period derivative \dotfill & $-8.37(3) \times 10^{-21}$ \\
Eccentricity Upper Limit \dotfill & 7 $\times 10^{-5}$\\
Mass Function (\msun) \dotfill & $0.0002321251(87)$ \\
Min. Companion Mass (\msun)\tnote{a} \dotfill & $\geq$\,0.0798\\
Med. Companion Mass (\msun)\tnote{b}\dotfill & 0.0927\\
\hline
\end{tabular}
\begin{tablenotes}
    \item[a] Assuming M$_{psr}$ = 1.4\msun\ and inclination $i = 90\degr$.
    \item[b] Assuming M$_{psr}$ = 1.4\msun\ and a random inclination distribution with $i = 60\degr$.
\end{tablenotes}
\end{threeparttable}
}
\end{table}

\subsection{Contributions to Observed Acceleration}\label{acc}
The observed spin-down \textit{$\dot{P}_{\textrm{obs}}$} is comprised of a combination of the intrinsic spin-down of the pulsar \textit{$\dot{P}_{int}$} and several external accelerations, such as the Shklovskii effect (a$_{\textrm{PM}}$, i.e.~the transverse Doppler effect from the pulsar's proper motion) \citep{shk70}, the cluster potential acceleration (a$_{\textrm{GC}}$)\citep{phi93}, the galactic potential acceleration (a$_{\textrm{G}}$)\cite[e.g.~][]{nic95}, and acceleration due to nearby objects (a$_{\textrm{NN}}$) \citep{dai23}.
\begin{equation}
    \left(\frac{\dot{P}}{P}\right)_{\textrm{obs}} = \left(\frac{\dot{P}}{P}\right)_{\textrm{int}} + \frac{a_{\textrm{GC}} + a_\textrm{G} + a_{\textrm{PM}} + a_{\textrm{NN}}}{c}.
\end{equation}
Typically, the cluster acceleration a$_\textrm{{GC}}$ dominates over the other terms, but without significant measurements on the binary non-Keplerian parameters (such as $\dot{P_{\textrm{orb}}}$), we can only make order-of-magnitude estimates of the contributions of each acceleration term \citep{fre17,pra17}. We estimate the acceleration due to the galactic potential acceleration to be $a_\textrm{G}/c$ $\approx$ $-1.3\times$10$^{-18}$\,s$^{-1}$, within a factor of $\sim$2 using Eq.~(3.5) from \cite{nic95}, due to limitations in our knowledge of the Galactic potential, with R$_0$ and $\Theta_0$ values reported in \citet{rei19} and cluster values from the Baumgardt catalog\footnote{\url{https://people.smp.uq.edu.au/HolgerBaumgardt/globular/parameter.html}} \citep{bau19}. For the Shklovskii effect \citep{shk70}, we estimate a$_{\textrm{PM}}$/c $\approx$ 6.0$\times$10$^{-22}$\,s$^{-1}$ and assuming that the pulsar has the same proper motion of the cluster. Lastly, we estimate $|a_{\textrm{NN}}/c|$ $\approx$ 1.5$\times$10$^{-19}$\,s$^{-1}$ using Eq.~(4) from \citet{dai23}. 

\begin{figure}
    \centering
    \includegraphics[width=\linewidth]{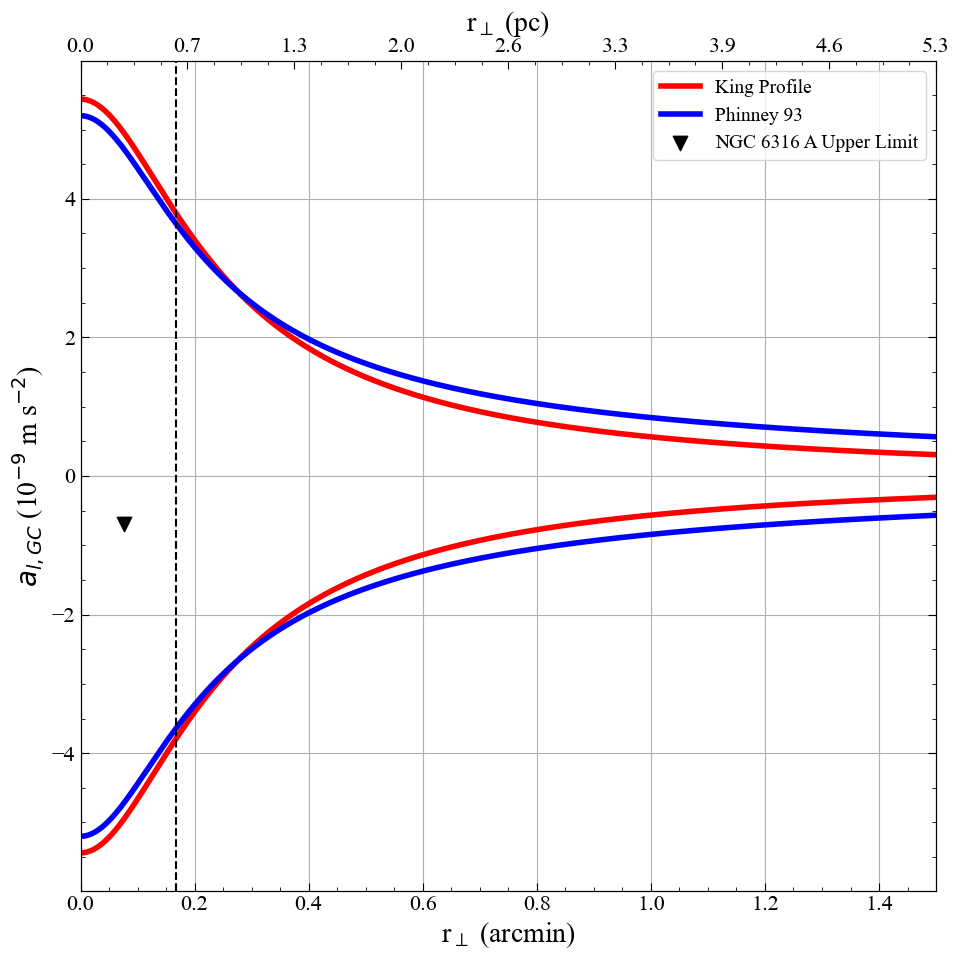}
    \caption{Maximum line of sight cluster acceleration profiles calculated using Eq.~\ref{phin3.5} and Eq.~\ref{freire} as a function of the angular offset from the center of NGC~6316. The dashed vertical line denotes the core radius. The upper limit for the cluster acceleration from PSR~J1716$-$2808A, assuming zero intrinsic spin-down, is shown via the inverted triangle and calculated via Eq.~\ref{ulim}. Given the estimated upper limit of PSR~J1716$-$2808A, we cannot constrain properties of the cluster without either detecting more MSPs or by measuring the orbital period derivative of PSR~J1716$-$2808A through continued timing.}
    \label{accel}
\end{figure}

\subsubsection{Acceleration Profiles}
We can derive an upper limit for the line-of-sight cluster acceleration, $a_{l,\textrm{GC}}/c \approx$ $-2.3\times10^{-18}$\,s$^{-1}$, by assuming zero intrinsic spin-down and subtracting the other acceleration terms from the observed acceleration $\dot{P}_{\textrm{obs}}/P$,
\begin{equation}
    \label{ulim}
    a_{l,\textrm{GC}} + \frac{\dot{P}_\textrm{{int}}}{P}c = \frac{\dot{P}_\textrm{{obs}}}{P}c - \mu^2d - a_{\textrm{G}} - a_\textrm{{NN}},
\end{equation}
which is plotted in Fig.~\ref{accel} as an inverted triangle for PSR~J1716$-$2808A. The solid lines represent the predicted maximum line-of-sight acceleration due to the cluster potential, calculated using two different models: Eq.~(3.5) from \cite{phi93}, shown via the blue curve in Fig~\ref{accel},
\begin{equation}
    \label{phin3.5}
	\frac{\textrm{max }a_{l,\textrm{GC}}}{c} = \frac{3 {\sigma_{v}}^2}{2c(r^2_c + r_{P}^2)^{1/2}},
\end{equation}
where $\sigma_{v}$ is the 1D central velocity dispersion \citep[$\sigma_v \approx$ 7.6 km s$^{-1}$,][]{der23}, r$_c$ is the cluster core radius (r$_c \sim 10\arcsec$), and r$_{P}$ is the projected distance from the pulsar to the cluster center (r$_P \sim 4.5\arcsec$), and an analytical model using the mass distribution presented in \citet{kin62} and described in \citet{fre05},\begin{equation}
    \label{freire}
    a_{l,\textrm{GC}}(x) = \frac{9 \sigma_{v}^2}{D\,r_c}\frac{l}{x^3}\left(\frac{x}{\sqrt{1+x^2}} - \textrm{sinh}^{-1}x \right),
\end{equation}
where \textit{D} is the distance \citep[D $\sim$11.3\,kpc,][]{der23}, \textit{l} is the distance in core radii to the plane of the sky that passes through the center of the cluster, so that $x = \sqrt{l^2 + x_{\perp}}$, and $x_{\perp} = r_{\perp}/r_c$, with $r_{\perp}$ characterized as a constant angular offset from the center. We maximize the value of \textit{l} at each $r_{\perp}$, in order to calculate the maximum values of $a_{l,\textrm{GC}}$, shown via the red curve in Fig~\ref{accel}. 

We can estimate the intrinsic spin-down acceleration ($\dot{P}_\textrm{{int}}$/\textit{P}) required in order to keep the predicted cluster acceleration as seen by the pulsar consistent with the maximum predicted lines (as shown in red and blue in Fig.~\ref{accel}). This intrinsic spin-down is $\approx 1.3\times10^{-17}$\,s$^{-1}$, which lets us place an upper limit of $\lesssim 3\times 10^8$\,G for the pulsar's magnetic field. Based on the acceleration estimates, the apparent spin up of the pulsar is due to a combination of the cluster and galactic potential acceleration. As the cluster is in an highly eccentric orbit confined within the galactic bulge, we expect a strong contribution from the galactic potential \citep{der23}. Continued timing of this MSP as well as discovery and timing of other MSPs within NGC~6316 will allow us to disentangle the effects of the cluster potential and galactic potential.  

\begin{figure}[t!]
    \centering
    \includegraphics[width=0.6\linewidth]{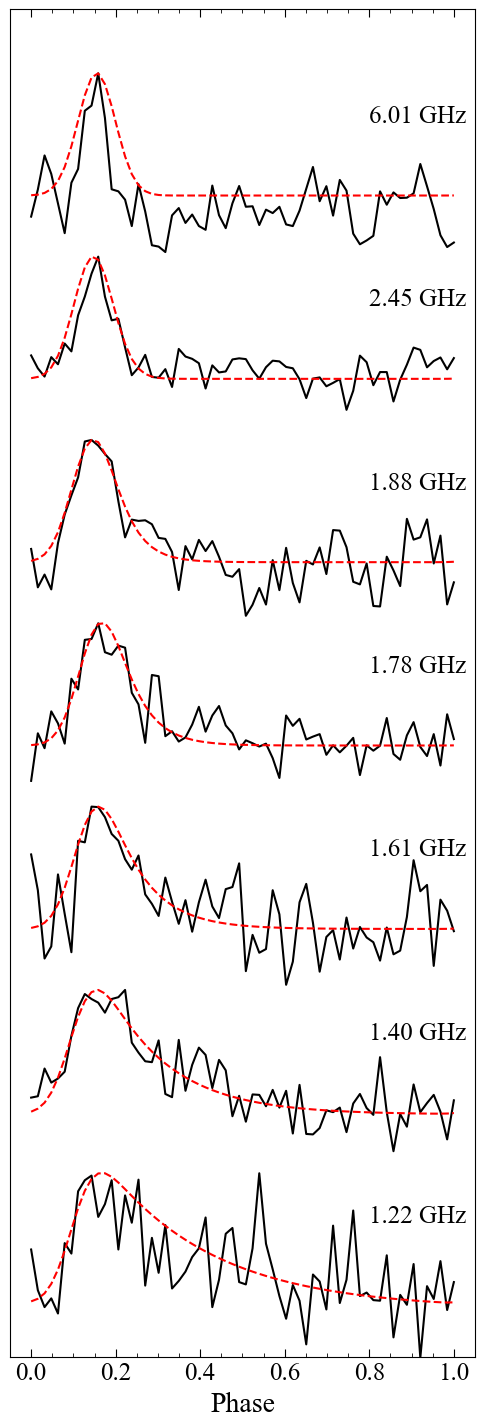}
    \caption{Estimated scattering profiles for the flux calibrated summed GBT L$-$band subbands (1.22, 1.40, 1.61, 1.78\,GHz), uncalibrated GBT S$-$band subbands (1.88, 2.45\,GHz) and GBT C$-$band observation (6.01\,GHz). The normalized subbands are plotted in black and the estimated scattering profile is plotted in red with: $\tau_{\textrm{1GHz}} \sim 1.4 \pm$ 0.1 ms.}
    \label{scatter}
\end{figure}
\subsection{Scattering and Flux Measurements}
We used {\tt PSRCHIVE} and our calibrated GBT L$-$band observations to measure PSR~J1716$-$2808A's flux density S$_\textrm{1.5\,GHz} \approx$ 43$\pm$0.8\,$\mu$Jy. We detected no measurable polarization and therefore could not constrain the rotation measure (RM). Due to our various calibrated and uncalibrated observations, we attempted to measure the pulse broadening from scattering through three approaches. We used the $\tau_s \propto \nu^{-4}$ relation to scale the profiles to 1\,GHz and convolved the profiles with a one-sided exponential function to estimate the scattering using MCMC to sample $\tau_s$, \citep[e.g.][]{cordes_space_1986, cordes_diffractive_1998, jing_fast_2025}. The reported errors are the statistical confidence intervals computed from the MCMC samples. First, we summed all the flux calibrated L$-$band observations using {\tt psradd}, aligning the observations using the summed profile as a template (with the {\tt -P} flag) with 6 subbands, measuring $\tau_{\textrm{1GHz}} \sim 1.41 \pm 0.09$\,ms. In our second attempt, we co-added our uncalibrated Parkes UWL observations, using the same alignment procedure to increase signal to noise and measure scattering across 8 subbands, measuring $\tau_{\textrm{1GHz}} \sim 1.70 \pm 0.09$\,ms. Additionally, we created subbanded profiles from the GBT S$-$band observations to measure scattering across the calibrated L$-$band and uncalibrated S$-$band observing bands, measuring $\tau_{\textrm{1GHz}} \sim 1.00 \pm 0.68$\,ms. Given the sensitivity of our flux calibrated GBT L$-$band observations, we estimate the scattering to be roughly $\tau_{\textrm{1GHz}} \sim 1.4 \pm 0.1$\,ms. Based on the systematic uncertainties due to low signal-to-noise and only minimal scattering at our observing frequencies (see Fig.~\ref{scatter}), the best way to more accurately measure the scattering would involve higher signal-to-noise observations with the new GBT Ultra$-$wideband Receiver (UWBR), which would also allow for more accurate flux density, spectral index, and potentially polarization measurements.

\section{Gamma Ray Pulsation Search}
We performed a search for gamma-ray pulsations from PSR~J1716$-$2808A in the \textit{Fermi}-LAT data. We selected Pass 8 {\tt SOURCE}-class photons \citep{atw13} detected by the \textit{Fermi}-LAT between 2008 August 5 and 2025 February 2 with an energy range of 0.1-5\,GeV photons, from within a 5\,deg radius around the radio position of the pulsar. Photon weights were calculated using the {\tt gtsrcprob} routine within {\tt Fermitools} \citep{fermitools} and the {\tt P8R3\_SOURCE\_V3} instrument response functions. These weights correspond to the probability of each photon being associated with the pulsar as opposed to a background source. {\tt gtsrcprob} uses a combination of a spatial model and a spectral flux model using the \textit{Fermi}-LAT 14-year Source Catalog (4FGL-DR4) to calculate the weights \citep{ker11,bru19,abd22,bal23}. 

We used the single photon timing code {\tt event\_optimize} within {\tt PINT} \citep{pint_ascl,pint} to search for pulsations. Given that external environmental acceleration causes observed higher order spin derivatives, we cannot accurately extrapolate a 3\,year timing solution to a 17\,year pulsation search without searching for those spin derivatives. As a starting point, we expect the strongest chance of detection in the photon data range corresponding to our radio timing baseline. We searched for any pulsations from MJD 59000 to MJD 60800 with no successful detection (H-test statistic of 1.22 or $\sim0.5\sigma$). This is likely due to the high background confusion given the cluster environment, and proximity to the Galactic plane and bulge. 

\section{Conclusions}
In this paper, we report the discovery and timing solution of the first pulsar in the GC NGC~6316, PSR~J1716$-$2808A. The system is a 2.45\,ms pulsar in a compact binary system with a $\sim$ 0.1\,M$_\odot$ companion. The spin period derivative indicates that unknown combinations of intrinsic spin-down, and accelerations due to the cluster potential, the Galactic potential are impacting our measurement. As we measure a negative period derivative, the acceleration experienced by the pulsar is likely dominated by the cluster and galactic potential and the pulsar is being accelerated towards us from behind the cluster. Continued timing of this pulsar should allow us to measure an orbital period derivative and then disentangle the effects of spin-down and the cluster acceleration \citep{pra17}. However, to constrain cluster properties using accelerations, we will need to detect more pulsars within the cluster. Given the mass (3.5$\times$10$^{5}$\,M$_\odot$ \citep{bau23}), the similarities to 47 Tuc, and the number of MSPs estimated via \gr\ spectral analysis \citep{dem19,wu21}, we predict that NGC~6316 houses many more MSPs to be discovered. Finding these MSPs will require more sensitive searches, and may be possible using uGMRT, MeerKAT or with the next generation of telescopes, such as the DSA, SKA, or ngVLA.

\begin{acknowledgements}
The National Radio Astronomy Observatory and Green Bank Observatory are facilities of the U.S. National Science Foundation operated under cooperative agreement by Associated Universities, Inc. Support for this work was provided by the NSF through the Grote Reber Fellowship Program administered by Associated Universities, Inc./National Radio Astronomy Observatory. The Parkes radio telescope (Murriyang) is part of the Australia Telescope National Facility which is funded by the Australian Government for operation as a National Facility managed by CSIRO. We acknowledge the Wiradjuri People as the traditional owners of the Observatory site. We acknowledge the Wallumedegal People of the Darug Nation and the Wurundjeri People of the Kulin Nation as the traditional owners of the land where this work was carried out. SMR is a CIFAR Fellow and is supported by the NSF Physics Frontiers Center award 2020265 and AAG award 2510064.
\end{acknowledgements}

\facilities{GBT, Parkes, \textit{Fermi}-LAT}

\software{PRESTO, PINT, FERMITOOLS, SPIDER\_TWISTER}

\bibliography{biblio}{}
\bibliographystyle{aasjournalv7}
\end{document}

%% file: macros.tex
\def\apjl{ApJ}
\def\apjs{ApJS}
\def\ao{Appl.~Opt.}
\def\aap{A\&A}